\documentclass[preprint,12pt]{elsarticle}

\usepackage{amssymb}
\usepackage{amsmath}
\usepackage{graphicx}
\usepackage{bm}
\usepackage{epsfig}
\usepackage{natbib}

\newcounter{bla}

\journal{Computer Physics Communications}

\begin{document}

\begin{frontmatter}



\title{CHIWEI: A code of goodness of fit tests for weighted and unweighted histograms}


\author[a]{N.D. Gagunashvili \corref{author}}

\cortext[author] {Corresponding author.\\\textit{E-mail address:} nikolai@simnet.is}
\address[a]{University of Akureyri, Borgir, v/Nordursl\'od, IS-600 Akureyri, Iceland}

\begin{abstract}
A  Fortran-77 program for  goodness of fit tests for  histograms with weighted entries as well as with unweighted entries is presented.  The code calculates test statistics for case of histogram with normalized weights of events and in case of unnormalized weights of events.
\end{abstract}


\end{frontmatter}



{\bf PROGRAM SUMMARY}
\vspace *{0.5 cm}

\begin{small}
\noindent
{\em Program Title:} CHIWEI                                        \\
{\em Journal Reference:}                                      \\
{\em Catalogue identifier:}                                   \\
{\em Licensing provisions:} none                                  \\
{\em Programming language:} Fortran-77                                  \\
{\em Computer:} Any Unix/Linux workstation or PC  with a Fortran-77 compiler                                              \\
{\em Classification:} 4.13, 11.9, 16.4, 19.4                   \\
{\em External routines/libraries used:} FPLSOR (M103) from CERN Program Library  \\
{\em Nature of problem:} The program calculates  goodness of fit test statistics for weighted histograms \\
{\em Solution method:}  Calculation of test statistics is done according  formulas presented in Ref. \cite{good} \\

\end{small}

\section{Introduction}

 A histogram with $m$ bins for a given probability density function  $p(x)$ is used to estimate the
 probabilities
\begin{equation}
p_i=\int_{S_i}p(x)dx, \; i=1,\ldots ,m \label{p1}
\end{equation}
 that a random event  belongs to  bin $i$. Integration in (\ref{p1}) is done over the bin  $S_i$.

A histogram can be obtained as a result of a random experiment with probability
 density function $p(x)$.
Let us denote the number of random events belonging to the $i$th bin
of the  histogram as $n_{i}$. The total number of events in the
histogram is equal to $n=\sum_{i=1}^{m}{n_i}$.  The quantity
$\hat{p}_i= n_{i}/n$ is an estimator
 of $p_i$  with expectation value $\textrm E \,\hat{p_i}=p_i$.

The problem of goodness of fit is to test the hypothesis
\begin{equation}
H_0: p_1=p_{10},\ldots, p_{m-1}=p_{m-1,0} \text{  vs.  } H_a: p_i \neq p_{i0} \text{  for some  } i,
\end{equation}
where $p_{i0}$ are specified probabilities, and $\sum_{i=1}^{m}
p_{i0}=1$. The test is used in a data  analysis for comparison
theoretical frequencies $np_{i0}$ with the observed frequencies
$n_i$.
The test statistic
\begin{equation}
X^2=\sum_{i=1}^{m} \frac{(n_i-np_{i0})^2}{np_{i0}} \label{basic}
\end{equation}
was suggested by Pearson \cite{pearson}. Pearson showed that the
statistic (\ref{basic}) has  approximately a $\chi^2_{m-1}$
distribution if the hypothesis
 $H_0$ is true.

To define a weighted histogram let us write the probability $p_i$
(\ref{p1}) for a given probability density function  $p(x)$  in the
form
\begin{equation}
p_i= \int_{S_i}p(x)dx = \int_{S_i}w(x)g(x)dx, \label{weightg}
\end{equation}
where
\begin{equation}
w(x)=p(x)/g(x) \label{fweight}
\end{equation}
 is the weight function and $g(x)$ is some other probability density function.
The function $g(x)$ must be $>0$ for points $x$, where $p(x)\neq 0$.
The weight $w(x)=0$ if $p(x)=0$, see Ref.  \cite{Sobol}.
Because of the condition $\sum_ip_i=1$ further we will call the above defined weights normalized weights as opposite to the unnormalized weights $\check{w}(x)$ which are $\check{w}(x)=const\cdot w(x)$.

The histogram with normalized weights
 was obtained from a random experiment with a probability density function $g(x)$, and the weights of the events were calculated according to (\ref{fweight}). Let us denote the total sum of the weights of the events in the $i$th bin of the histogram as
\begin{equation}
W_i= \sum_{k=1}^{n_i}w_i(k) \label{ffweight}
\end{equation}
and the total sum of squares of weights as
\begin{equation}
W_{2i}= \sum_{k=1}^{n_i}w_i(k)^2,
\end{equation}
where $n_i$ is the number of events at bin $i$ and $w_i(k)$ is the weight of the $k$th event in the $i$th bin. The total number of events in the histogram is equal to $n=\sum_{i=1}^{m}{n_i}$, where $m$ is the number of bins. The quantity $\hat{p}_i= W_{i}/n$ for the histogram with normalized weights is the estimator of $p_i$ with the expectation value $\textrm E \,
\hat{p_i}=p_i$. Note that in the case where $g(x)=p(x)$, the weights of the events are equal to 1 and the histogram with normalized weights
is the usual histogram with unweighted entries.

 For weighted histograms again the problem of goodness of fit is to test the hypothesis
\begin{equation}
H_0: p_1=p_{10},\ldots ,p_{m-1}=p_{m-1,0} \text{  vs.  } H_a: p_i \neq p_{i0} \text{  for some  } i, \label{hipoth}
\end{equation}
where $p_{i0}$ are specified probabilities, and $\sum_{i=1}^{m} p_{i0}=1$.

The test statistic that is a generalization of Pearson's statistic (\ref {basic}) was proposed in \cite{goodness} for cases of histograms with normalized weights of entries as well as with unnormalised weights of entries.  A code for the calculation of test statistics   is presented in this article.
As shown in \cite{goodness} if hypothesis $H_0$ (\ref{hipoth}) is  true then the statistic for a histogram with normalized weighted entries  has approximately the $\chi^2_{m-1}$ distribution and for a histogram with unnormalized weighted entries has $\chi^2_{m-2}$ distribution.

Use of the proposed test is inappropriate if any expected
count in bin of histogram is below 1 or if the expected count is less than 5 in more than 20\% of the bins. This empirical restriction known for the
usual chi-square test  \cite{moore} is quite reasonable for weighted histograms.

\textbf{Information for readers.} Recently, another paper dedicated to weighted histograms has been published in "Computer Physics Communication``, see Ref. \cite{chicom}. The same author has presented a program for calculating test statistics to compare weighted histogram with unweighted histogram and two histograms with weighted entries. The test can be used for the comparison of experimental data distributions with simulated data distributions as well as for the two simulated data distributions.

\section{Computer program}

CHIWEI is  subroutine which can be called from Fortran program for the calculation of test statistics.\\

\noindent
{\bf Usage}\\

\noindent
\verb"CALL CHIWEI(P,W1,W2,N,NCHA,MODE,STAT,NDF,IFAIL)"\\

\noindent
{\bf {\it {Input Data}}}\\

\noindent
P -- one dimensional real array of probabilities $p_i$ \\

\noindent
W1 -- one dimensional array, sum of weights $W_i$ in each bin \\

\noindent
 W2 -- one dimensional array,  sum of squares of weights $W_{2i}$ in each bin\\

\noindent
N -- number of events $n$\\

\noindent
NCHA -- number of bins $m$ \\

\noindent
 MODE -- must be equal to 1 for a histogram with normalized weights, and equal 2 for histogram with unnormalized weights\\
\newpage
\noindent
{\bf {\it {Output data}}}\\

\noindent
STAT -- test statistic following a chi-square distribution with NDF degrees of freedom if hypothesis $H_0$ is true  \\

\noindent
NDF -- number of degree of freedom (will be $m$-MODE)\\

\noindent
IFAIL -- will be $>0$ if calculation is not successful.

\section{Test run}

We take a distribution
\begin{equation}
p(x)\propto \frac{2}{(x-10)^2+1}+\frac{1}{(x-14)^2+1} \label{weight}
\end{equation}
 defined on the interval $[4,16]$ and representing two so-called Breit-Wigner peaks. Two cases of the probability density function $g(x)$ are considered

\begin{equation}
g_1(x)=p(x)   \label{prc}
\end{equation}

\begin{equation}
g_2(x)\propto\frac{2}{(x-9)^2+1}+\frac{2}{(x-15)^2+1} \label{real}
\end{equation}

 Distribution (\ref{prc}) gives an unweighted histogram and the method coincides with  Pearson's chi square test.
 Distribution (\ref{real}) has the same form of parametrization as (\ref{weight}), but with different values of the parameters.
 Three cases of histograms were considered: unweighted histogram, histogram with weights $p(x)/g_2(x)$ and histogram with unnormalized weights $2p(x)/g_2(x)$. Histograms with 5 bins were created  by simulation 1000 entries  for each case. The results of the calculations are presented below. Program PROB(G100) \cite{cern} has been used for calculating p-values.  \\

 \textbf{Test 1 }

\verb"            INPUT"\\

\noindent
\verb"P        0.0296    0.1106    0.4460    0.2067    0.2072"\\
\verb"W1      26.0000  115.0000  454.0000  183.0000  222.0000"\\
\verb"W2      26.0000  115.0000  454.0000  183.0000  222.0000"\\
\verb"N     1000"\\
\verb"NCHA     5"\\
\verb"MODE     1"\\

\verb"            OUTPUT"\\

\noindent
\verb"STAT     4.5291"   \hspace *{3cm}       (p-value=0.3391)          \\
\verb"NDF      4"\\
\verb"IFAIL    0"\\

\textbf{Test 2 }

\verb"            INPUT"\\

\noindent
\verb"P        0.0296    0.1106    0.4460    0.2067    0.2072"\\
\verb"W1      36.0112  106.1355  458.3037  197.8123  205.7211"\\
\verb"W2      28.2698   56.9601  938.7897  363.4649  172.2003"\\
\verb"N     1000"\\
\verb"NCHA     5"\\
\verb"MODE     1"\\

\verb"             OUTPUT"\\

\noindent
\verb"STAT     2.3380"         \hspace *{3cm}       (p-value=0.6738)\\
\verb"NDF      4"\\
\verb"IFAIL    0"\\

\textbf{Test 3 }

\verb"             INPUT"\\

\noindent
\verb"P        0.0296    0.1106    0.4460    0.2067    0.2072"\\
\verb"W1      72.0225  212.2710  916.6075  395.6246  411.4423"\\
\verb"W2     113.0790  227.8403 3755.1587 1453.8595  688.8014"\\
\verb"N     1000"\\
\verb"NCHA     5"\\
\verb"MODE     2"\\

\verb"            OUTPUT"\\

\noindent
\verb"STAT     2.2398"      \hspace *{3cm}       (p-value=0.5241)\\
\verb"NDF      3"\\
\verb"IFAIL    0"\\

\end{document}